\documentclass[a4paper,11pt]{article}
\pdfoutput=1 

\usepackage{jheppub} 

\usepackage[T1]{fontenc} 

\usepackage{amsmath, amsthm, mathtools}
\usepackage{cancel}
\usepackage{graphicx}
\usepackage{subfig}
\usepackage{xspace}

\usepackage[toc,page]{appendix}

\newcommand{\PZ}{\ensuremath{\text{Z}}\xspace}
\newcommand{\PZprime}{\ensuremath{{\PZ}^{\prime}}\xspace}
\newcommand{\mZprime}{\ensuremath{m_{\PZprime}}\xspace}
\newcommand{\mDark}{\ensuremath{m_{\text{dark}}}\xspace}
\newcommand{\aDark}{\ensuremath{\alpha_{\text{dark}}}\xspace}
\newcommand{\rinv}{\ensuremath{r_{\text{inv}}}\xspace}
\newcommand{\ptD}{\ensuremath{p_{T}D}\xspace}
\newcommand{\pt}{\ensuremath{p_{T}}\xspace}
\newcommand{\met}{\ensuremath{\cancel{E}_{T}}\xspace}
\newcommand{\mt}{\ensuremath{m_{T}}\xspace}
\newcommand{\mj}{\ensuremath{m_{j}}\xspace}
\newcommand{\PYTHIA}{\texttt{PYTHIA8}\xspace}
\newcommand{\DELPHES}{\texttt{DELPHES}\xspace}

\title{Autoencoders for Semivisible Jet Detection}

\author[a]{Florencia Canelli}
\author[b]{Annapaola de Cosa}
\author[c]{Luc Le Pottier}
\author[b]{Jeremi Niedziela}
\author[d]{Kevin Pedro}
\author[e]{Maurizio Pierini}

\affiliation[a]{University of Zurich, Switzerland}
\affiliation[b]{ETH Zurich, Switzerland}
\affiliation[c]{University of California, Berkeley, USA}
\affiliation[d]{Fermi National Accelerator Laboratory, Batavia, IL 60510, USA}
\affiliation[e]{European Organization for Nuclear Research (CERN), Switzerland}

\emailAdd{florencia.canelli@physik.uzh.ch}
\emailAdd{adecosa@phys.ethz.ch}
\emailAdd{luclepot@berkeley.edu}
\emailAdd{jeremi.niedziela@cern.ch}
\emailAdd{pedrok@fnal.gov}
\emailAdd{maurizio.pierini@cern.ch}

\abstract{

The production of dark matter particles from confining dark sectors may lead to many novel experimental signatures. Depending on the details of the theory, dark quark production in proton-proton collisions could result in semivisible jets of particles: collimated sprays of dark hadrons of which only some are detectable by particle collider experiments. The experimental signature is characterised by the presence of reconstructed missing momentum collinear with the visible components of the jets. This complex topology is sensitive to detector inefficiencies and mis-reconstruction that generate artificial missing momentum. With this work, we propose a signal-agnostic strategy to reject ordinary jets and identify semivisible jets via anomaly detection techniques. A deep neural autoencoder network with jet substructure variables as input proves highly useful for analyzing anomalous jets.
The study focuses on the semivisible jet signature; however, the technique can apply to any new physics model that predicts signatures with anomalous jets from non-SM particles. 
}

\begin{document} 
\maketitle
\flushbottom

%
\section{Introduction}
\label{sec:intro}

Hidden Valley models of dark matter allow for a strong interaction in the dark sector \cite{hidden_valley}. Such a strong interaction would cause showering of the dark particles, producing both invisible dark matter and Standard Model (SM) hadrons. As a result, so-called semivisible (SV) jets would be formed, containing some fraction $r_{inv}$ of invisible particles \cite{sv_jets, Cohen:2017pzm}. Depending on the mass of the mediator \mZprime, SV jets could be produced at the Large Hadron Collider (LHC) energies and registered in detectors \cite{cmscollaboration2021search}. 

Events containing such jets would be characterised by missing transverse energy \met being aligned with a jet, which is a scenario difficult to distinguish from detector issues and therefore usually neglected in LHC data analyses. Moreover, the details of the Hidden Valley physics depend on a number of parameters. There are four parameters that most influence the kinematic behavior in the final state:

\begin{itemize}
\item \aDark -- the coupling constant of the dark sector strong interaction equivalent,
\item \mZprime -- the mass of the mediator,
\item \mDark -- the mass of the dark hadrons,
\item \rinv -- the fraction of stable, invisible dark hadrons.
\end{itemize}

These parameters could be a function of other parameters in a fully-specified Hidden Valley model, as well as parameters that impact the modeling of strong dynamics in the event generator (hadronization, fragmentation). This leads to an extremely large model space with a huge number of possible scenarios that can easily evade any constraints from e.g. cosmological measurements. It is clearly impractical to perform dedicated searches for all possible model variations.

In order to gain robustness against detector effects, as well as details of the model implementation, anomaly detection techniques such as unsupervised deep learning can be used. Those are designed to detect objects significantly different from the training sample, without prior knowledge of signal characteristics. The presented approach is suitable to search for a wide class of new physics models containing anomalous jets, including the SV jets stemming from Hidden Valley models, which are the main focus of this work.

Here, an autoencoder-based analysis for SV jets detection will be presented. Unlike previous attempts \cite{qcd_or_what, new_physics_deep_autoencoders, lhc_olympics, dark_machines, cheng2021variational, olmo} that use jet images (four-momenta of jet constituents), we incorporate high-level jet features and substructure variables computed based on the four-momenta of the jet constituents: Energy Flow Polynomials (EFPs)~\cite{eflow}, Energy Correlation Functions (ECFs) \cite{ecf} and their ratios: C\textsubscript{2} and D\textsubscript{2}, as well as the jet \pt dispersion \ptD~\cite{ptd} and jet axes \cite{jets_at_lhc}. These prove to be highly useful in discriminating between the QCD background and the SV jet signal \cite{svj_substructure}.

The trained model can then be used to tag anomalous jets, as the autoencoder has been trained to recognize QCD jets in the signal region, but not SV signal jets (or any other anomalous jets, such as emerging jets \cite{emerging_jets}).

It is worth noting that, while the performance is assessed using SV jet models as a benchmark, the training process is independent of the signal implementation, unlike in case of the supervised learning \cite{graph_net_dark_showers}. The aforementioned choice of substructure variables may have an impact on sensitivity to different anomalous jet models. However, selected features capture various properties of jet substructure, such as number of sub-jets, jet width or distribution of momenta among constituents, therefore making this approach suitable for a wide range of new physics models. Given that the landscape of possible Hidden Valley scenarios is so vast, using one classifier (even parametric \cite{parametric_ml}) per signal model becomes impractical.

%
\section{Data samples}
\label{sec:data}

\subsection{Generation}
Both multi-jet QCD and dark sector samples are generated using \PYTHIA~\cite{pythia} with model parameters identical as in \cite{hepdata.115426} and reconstructed with \DELPHES~\cite{delphes} providing a detector response approximating that of CMS. Proton-proton collisions at a centre-of-mass energy of 13 TeV are considered, with approximately 50 collisions per bunch crossing. The particle flow reconstruction algorithm distributed with \DELPHES is used. In particular, jets are clustered from reconstructed particles and pileup reduction is applied.

To examine the robustness of autoencoder models to varying signal parameters, multiple SV jet samples were generated with values of the fraction of stable jet hadrons $\rinv \in \{0.3, 0.5, 0.7\}$ and the \PZprime boson mass $\mZprime \in \{1.5, 2.0, 2.5, 3.0, 3.5, 4.0 \}$. The number of events generated varies depending on the sample, accounting for the different selection efficiency (see Sec.~\ref{subsec:preselection}):

\begin{itemize}
    \item QCD: $3.7 \cdot 10^7$ events,
    \item SV jets, $\mZprime > 1.5$ TeV: 50k events per sample,
    \item SV jets, $\mZprime = 1.5$ TeV, $\rinv \in \{0.3, 0.5\}$: 100k events per sample,
    \item SV jets, $\mZprime = 1.5$ TeV, $\rinv = 0.7$: 140k events per sample.
\end{itemize}

\subsection{Preselection}
\label{subsec:preselection}

A set of preselection criteria (based on those outlined in Ref.~\cite{sv_jets}) is applied to both training and testing samples on an event-by-event basis. These requirements are aimed at isolating the SV jet signal, as well as reproducing the realistic triggering capabilities of LHC experiments. The signal region is defined to include events passing the following criteria:
\begin{itemize}
    \item at least 2 jets with $|\eta| < 2.4$ and $\pt > 200$ GeV
    \item then, for the two leading jets:
    \begin{itemize}
        \item $|\Delta\eta| < 1.5$,
        \item $\mt > 1500$ GeV,
        \item $\met/\mt > 0.25$,
    \end{itemize}    
\end{itemize}
where $\eta$ is the pseudorapidity, \pt is the transverse momentum for each jet, $\Delta\eta$ is the absolute dijet $\eta$ difference, \met is the missing momentum in the transverse plane for the event, and \mt is the transverse mass of the leading dijet system and the \met.

The selection efficiency was found to be at the level of $0.13 \%$ for the QCD events and between $0.2$ and $15.3 \%$ for SV jets, depending on the signal model parameters. The number of QCD jets after applying selections is $\approx$100k (split between training, validation, and testing) and varies between 500 and 15k for SV jets, depending on the model parameters.

\subsection{Feature Selection}

Since the goal of this study is to find anomalies on the basis of tagging individual anomalous jets rather than anomalous events, a set of jet-level and jet substructure variables is determined for the training.

The coordinates of each jet ($\eta$ and $\phi$) are included in the training to allow the network to learn about problematic regions of the detector and avoid tagging noise or other detector failures as anomalous signals.

In order to avoid bias, the transverse momentum of the jet was not included in the training. Moreover, the jet \pt distribution was flattened (by applying appropriate weights) for the training.

A number of other jet-level and substructure variables have been considered for the training. A fraction of them were rejected due to poor signal-background discrimination. The correlations between remaining variables were tested, as autoencoders tend to perform better with uncorrelated inputs. It was found that for QCD and SV jets, the EFPs are fully correlated with each other, as well as with the girth \cite{jet_substructure}, Les-Houches Angularity (LHA) \cite{lha}, and normalized ECFs: e\textsubscript{2} and e\textsubscript{3}. Because of that, it was decided to only keep one of those variables for the training, namely EFP\textsubscript{1} \cite{eflow} which exploits pairwise relations between jet constituents.

The jet constituent four-momenta have also been considered; however, they proved not to bring any significant gain in performance, while increasing the complexity (${\approx}100$ instead of ${\approx}10$ input features) and training time of the autoencoder. This can be understood since EFPs and ECFs are included in the training and provide a linear basis for characterizing jet constituent substructure, making up for the discrimination that is lost when using high-level jet features rather than jet constituents.

\begin{figure}
    \centering
    \includegraphics[width=1.0\textwidth]{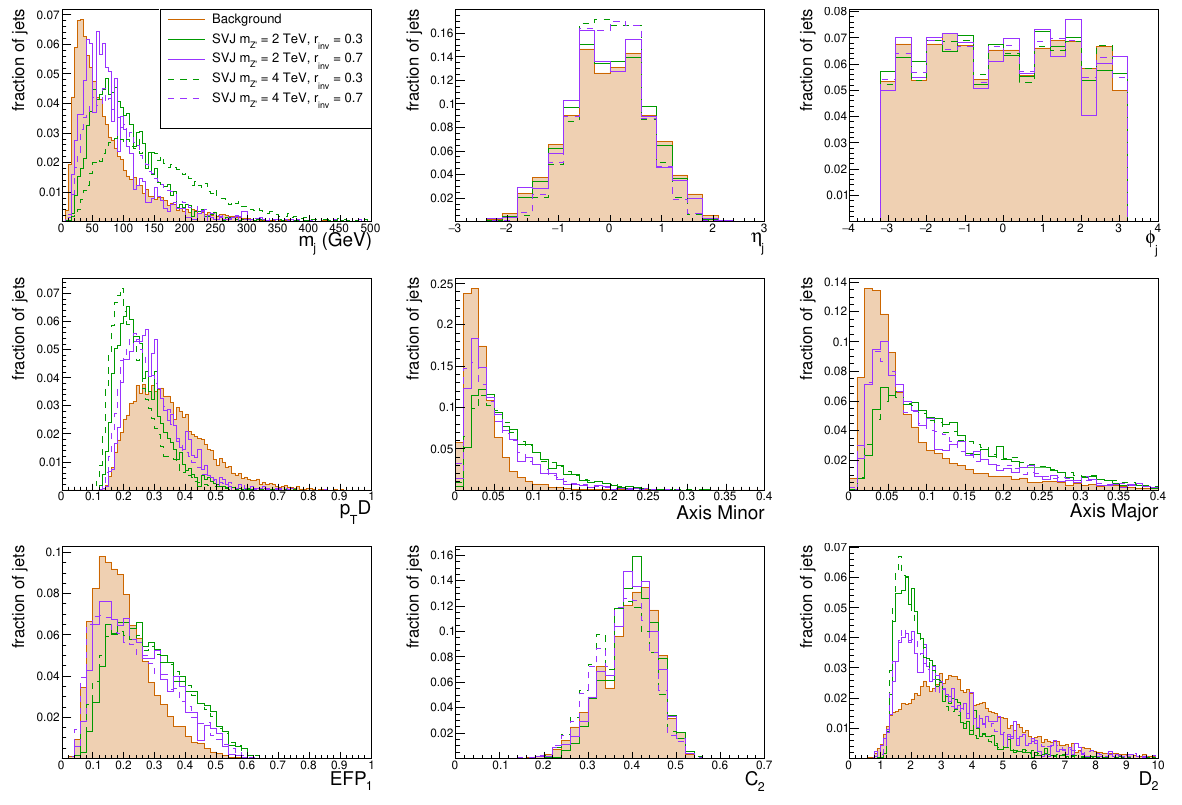}
    \caption{Distributions of input variables for QCD background and selected signal models.}
    \label{fig:inputs}
\end{figure}

Ultimately, the following set of jet-related input variables is selected: $\eta$, $\phi$, invariant mass \mj, jet fragmentation function \ptD, jet ellipse minor and major axes, EFP\textsubscript{1}, and ECF ratios: C\textsubscript{2} and D\textsubscript{2}. The distributions of those variables for the QCD background and selected signals is presented in Fig.~\ref{fig:inputs}.

The first three of these features are provided directly by the \DELPHES reconstruction algorithm. The remaining ones are calculated using a list of jet constituents, which consists of all particles in the event which lie within a cone of radius $R = 0.8$ from the jet axis.

All of the selected high level features discussed here provide some meaningful discrimination power, with the exception of the angular coordinates, $\eta$ and $\phi$. Since those are randomly distributed, they provide no meaningful information about QCD jet structure. However, given the ability of autoencoders to learn a compressed representation of input signals, it is feasible that they might also learn the locations and behaviors of defective detector components. For instance, they could learn location of energy spikes in certain detector locations as normal behavior that does not indicate an anomaly associated with a potential signal. Such detector malfunctions are not included in the \DELPHES reconstruction; however, they may be present in the collision data, on which the autoencoder can be trained.

%
\section{Models}\label{sec:models}

All unsupervised models presented in the following sections are trained using the QCD data sample described in Sec.~\ref{sec:data} as background, with a 70:15:15 split between training, validation, and testing, respectively.

\subsection{Autoencoders}\label{sec:models:ae}

An autoencoder is a neural network which approximates the identity transformation as the application of a function $f$ from the input space to a latent space (encoding) and a function $f^{-1}$ from the latent space back to the input space (decoding), which provides an approximation of the input as the output. The constraint forces the autoencoder to find hidden relationships between features in the input data set. In the case of a neural autoencoder, this constraint takes the form of a bottleneck in the network architecture: an input vector $\vec{x} \in \mathbb{R}^n$ is compressed into a latent space vector $\vec{l} \in \mathbb{R}^m$ with $m < n$, enforced by an $m$-node layer within the architecture of an autoencoder with input dimension $n$. This latent space vector is then reconstructed by a decoding network in an attempt to match the input values.

\begin{figure}\centering
\includegraphics[scale=0.5]{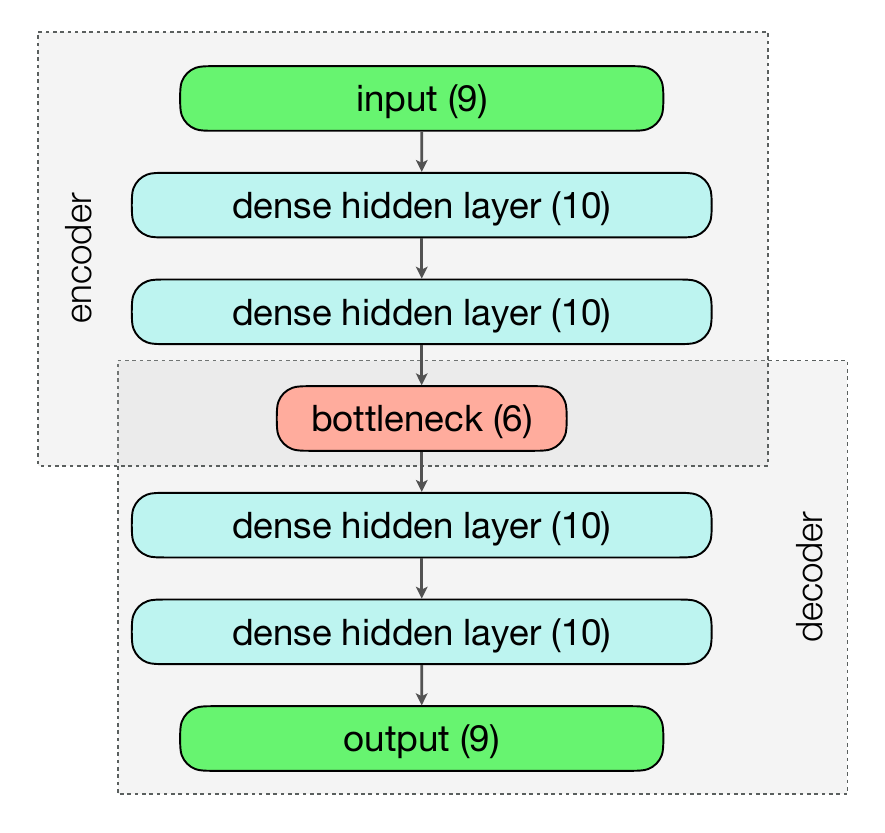}
\caption{Neural Network architecture.}
\label{nn_arch}
\end{figure}

The autoencoder is trained to minimize the reconstruction error of each sample, defined for an $n$-dimensional input feature vector $\vec{x}$ with components $x_1, ..., x_n$. A number of loss functions were considered and the mean absolute error (MAE) was selected, as it had the best performance on the considered benchmark data sets:
\begin{equation}
\label{eq:mae}
    \text{MAE} = \frac{1}{N}\sum_i^N \left|x^\prime_i - x_i\right|
\end{equation}

\noindent where $x_i$ is the original value and $x^\prime_i$ is the reconstructed value for the $i^{th}$ feature of a jet. The default choice of the loss function for autoencoders is often the mean squared error, which in our case indicated around 7\% decrease in performance compared to MAE. This can be understood as MAE tends to be more robust against outliers and most input variables have non-gaussian distributions with long tails (see Fig.~\ref{fig:inputs}).

The training data consists of the simulated QCD background events that fall within our signal region as per the cuts described in Sec.~\ref{sec:data}. 

Once a model is trained to recognize the background, one can evaluate it on unlabeled data and use the reconstruction errors as a measure of similarity of a given sample to the training data set. Typically, a model trained on QCD will return low reconstruction loss for samples similar to those it was trained on, and high loss for samples that are more complex \cite{ae_anomaly_hep,latent_space_classifiers}. In case of a search for new physics, an autoencoder would be trained and evaluated, with the largest 10\%, 1\%, and 0.1\% of its reconstruction losses isolated and analyzed.

\subsection{Alternative Models}
\label{sec:models:others}

Along with neural autoencoders - the simplest form of autoencoder - other types of autoencoders were considered: 
\begin{itemize}
    \item Variational Autoencoders~\cite{vae} (VAE): provide probabilistic representation of the input data, which can give higher performance in some applications.
    \item Principal Component Analysis (PCA) \cite{pca}: can be viewed as a linear autoencoder with no hidden layers, which allows determining the weights analytically.
\end{itemize}

The results obtained with the VAE and PCA will be reported in this work.
Sparse Autoencoders (SAE)~\cite{sae:eth} were investigated in the early stages of this study and found to give worse results than the dense autoencoder; therefore, they will not be included.

\subsection{Model Optimization}

In order to determine the optimal architecture, normalization, and hyper-parameters for the SV jet search, a large number of models were trained and evaluated. The optimization was focused on maximizing the Area Under Receiver Operating Characteristic (ROC) Curve (AUC). For each set of hyper-parameters and each choice of architecture, 80\% of best models were selected and the average AUC over all signal samples and all models was used to choose the highest performing one.

It was found that the best results are obtained by standardizing the data, i.e., by normalizing them such that the mean of the distribution is at zero and the standard deviation is unity. An extensive architecture scan was performed and the architecture with two hidden layers of size 10, followed by the bottleneck of size 6 was found to be optimal. The activation function of the hidden layers and the bottleneck is `elu' \cite{elu} and the output layer uses linear activation. 

The autoencoder models were implemented in Keras/TensorFlow ~\cite{keras,tensorflow}. Each model was trained for 200 epochs (with early stopping enabled), with a batch size of 256. The learning rate was set to $10^{-6}$ and the optimizer found to yield the best performance was `Nadam'~\cite{nadam}. 

\section{Results}

In this section, the results from an average autoencoder will be presented (unless stated otherwise). The average autoencoder was found by training 150 models with the optimal hyper-parameters and varying random seed. We select a model closest to the mean $\mu$ of AUC distribution (averaged out over all signal scenarios for given autoencoder model) and use models closest to $\mu\pm\sigma$, where $\sigma$ is the standard deviation, to estimate the uncertainty. No signs of overfitting were observed, and the loss evolution was smooth and saturated after around 100 epochs.

\subsection{Reconstructed Distributions}

\begin{figure}\centering
\includegraphics[width=1.0\linewidth]{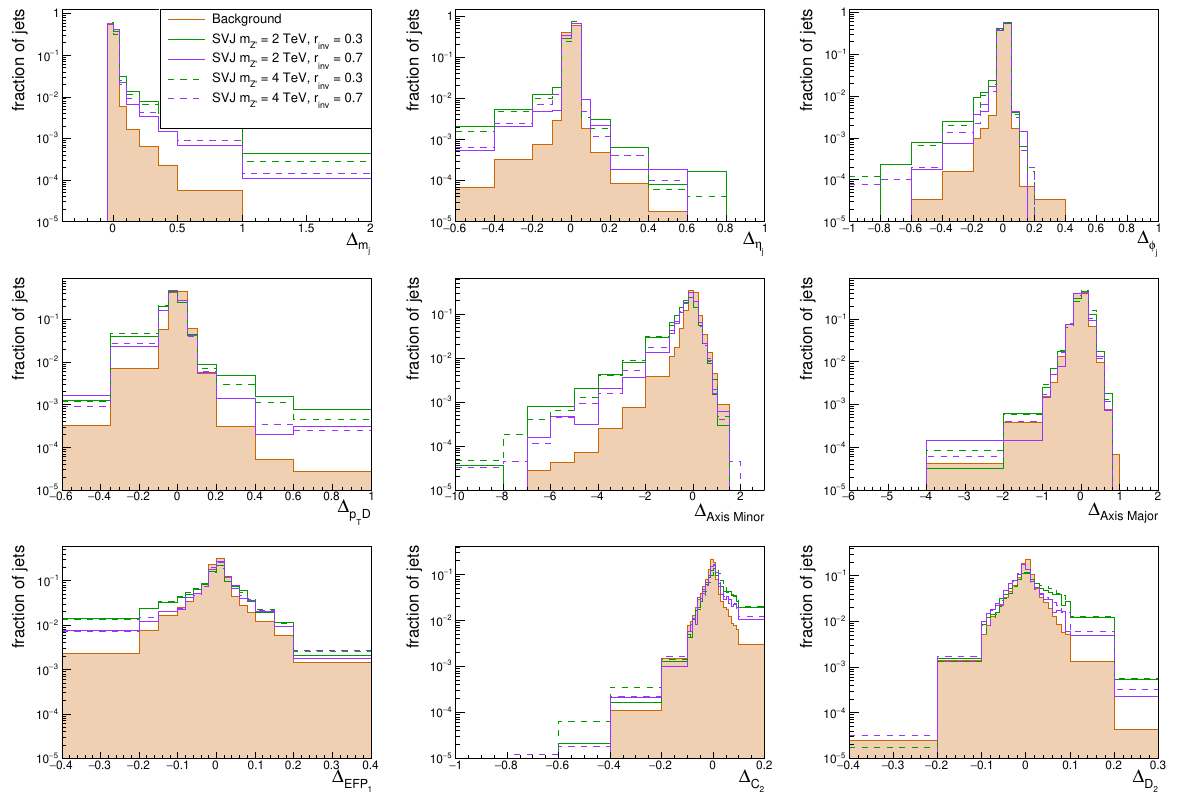}
\caption{Distributions of difference between input and reconstructed jet features for QCD and a few example SVJ signals.}
\label{fig:input_and_reco}
\end{figure}

The differences between input and reconstructed jet features are shown for both SVJ and QCD data in Fig.~\ref{fig:input_and_reco}. One can observe that, in almost all cases, the difference is larger for the signals than for the background. What is more, the distributions for the signal do not change with varying mass; however there is a slight difference in shape when moving from \rinv = 0.3 to 0.7.

\subsection{Reconstruction Loss}

\begin{figure}\centering
\includegraphics[width=1.0\textwidth]{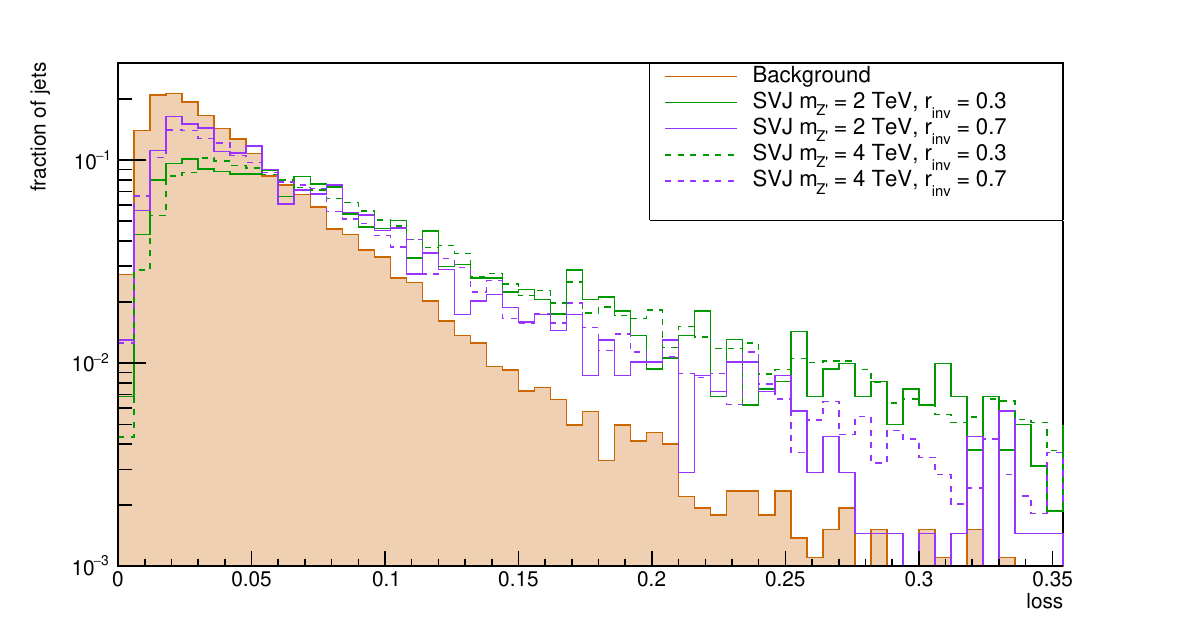}
\caption{Reconstruction loss for QCD and different SVJ signals.}
\label{fig:losses}
\end{figure}

An important figure of merit is the distribution of losses for the background and signal samples. As shown in Fig.~\ref{fig:losses}, QCD tends to have lower loss values than SVJ, providing good discrimination.

In order to achieve the best possible performance, alternative approaches to autoencoder training and evaluation were studied. For those, reference distributions were built for each node of the network from the training data set, using the reconstructed variables, MAE values, or even values from the latent space nodes. The performance was then evaluated by comparing values from the tested jets to those reference distributions. 

The comparison was done using the negative log-likelihood, as well as the Mahalanobis distance \cite{mahalanobis_distance}. It was found that none of the six combinations performed better than the MAE loss function.

\clearpage
\newpage

\subsection{Discrimination Power}
    
In order to compare discrimination power of different architectures, one can examine the ROC curves and AUC values for each generated signal. As a reference for the best-case performance, we considered a Boosted Decision Trees (BDT), providing a strong baseline for supervised algorithms. The BDT was implemented using SKLearn's AdaBoostClassifier, trained on a mixture of all signals, and had its hyper-parameters optimized for maximum AUC. The input features were the same as for the autoencoder. For the final comparison, 10 models were trained and the best one was selected (although the performance spread between different models was negligible).

\begin{figure}\centering
\includegraphics[width=1.0\textwidth]{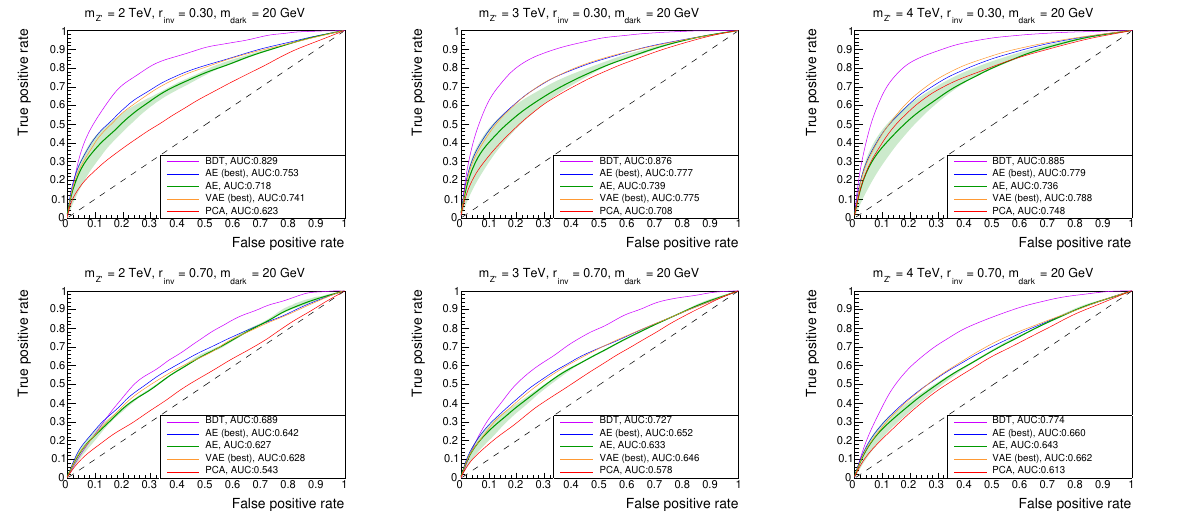}
\caption{Comparison of the ROC curves for the autoencoder (mean and best models), PCA, BDT and VAE (the best model out of 15 trained) for a few selected signals.}
\label{fig:rocs}
\end{figure}

A comparison of ROC curves for the autoencoder, BDT, PCA and VAE is shown in Fig. \ref{fig:rocs}. As can be seen, the BDT classifier provides the best performance regardless of the signal considered, although it is rather closely followed by the autoencoder. What is more, as will be shown in the further part of this section, the AE can become more powerful than the BDT if the latter was trained on a wrong signal hypothesis. The VAE has been evaluated on MAE reconstruction loss exclusively, neglecting the KL divergence \cite{kl_divergence}. The coefficient determining proportions of reconstruction and KL loss during the training has been scanned and optimized for maximum performance. It was found optimal to completely exclude KL loss during training, which leads to variance collapse to zero, showing that VAE tested on reconstruction loss only does not improve the results with respect to a regular AE. The PCA performance is significantly worse than those of autoencoders and BDT.

Figure \ref{fig:aucs} presents a comparison of the AUC values between the autoencoder and BDT for each of the signals. It is clear from this plot that signals with lower values of \rinv are much more efficiently identified, although all signals can be tagged at relatively high rates. Moreover, even though the autoencoder performs worse than the BDT in all signal cases, it is still quite good, with the added benefit of requiring no particular signal to train against.

\subsection{Robustness}

\begin{figure}\centering
\includegraphics[width=1.0\textwidth]{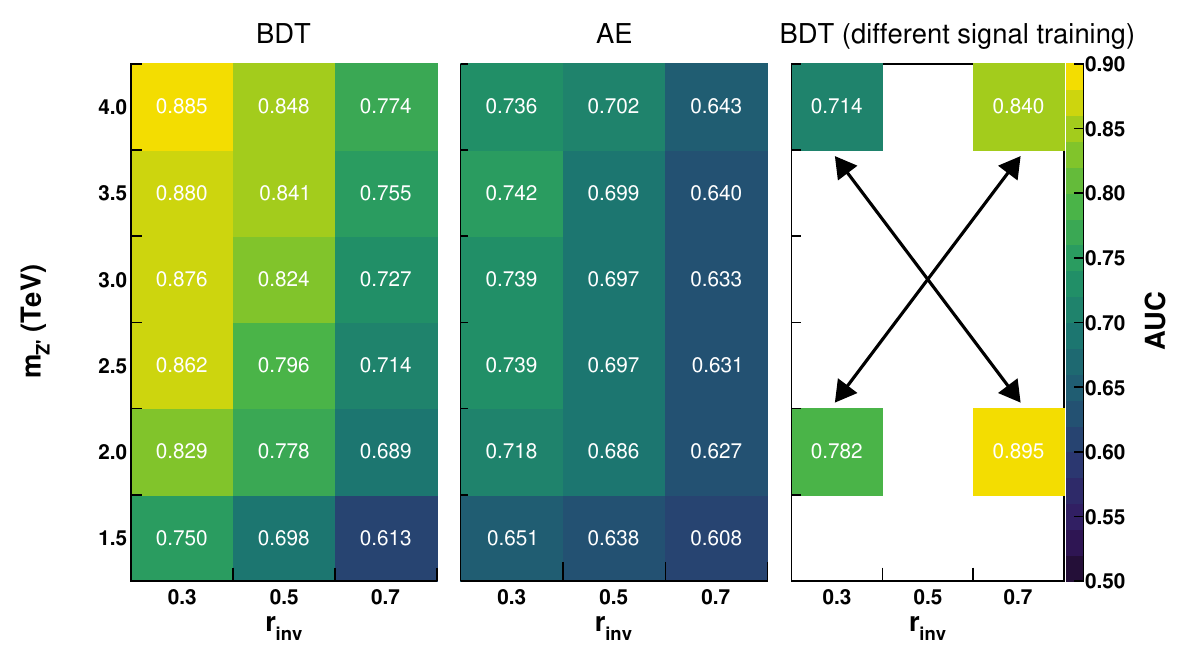}
\caption{Left and middle panels: comparison of AUC values of the autoencoder and a BDT trained on a mixture of all signal models with $\mDark = 20$~GeV. Right panel: AUC values for a BDT trained on a signal with parameters different from those it was tested on. E.g., the AUC value presented in the top left corner of the table comes from a model trained on the lower right corner sample.}
\label{fig:aucs}
\end{figure}

As an additional test, the autoencoder performance was compared to that of a BDT trained on the wrong signal hypothesis. This allows to study the generalization power of the autoencoder against that of a supervised model and assess their ability to discover ``by accident'' a signal other than the targeted one. As can be seen in the rightmost panel of Fig.\ref{fig:aucs}, a clear decrease in the performance can be observed in two cases, up to $19\%$ worse when trained on $\mZprime = 2$ TeV, $\rinv = 0.7$, but tested on $\mZprime = 4$ TeV, $\rinv = 0.3$. What is remarkable is that in this particular case, the BDT performs worse than the autoencoder, which was trained without any information about the signal characteristics.

\begin{figure}\centering
\includegraphics[width=0.6\textwidth]{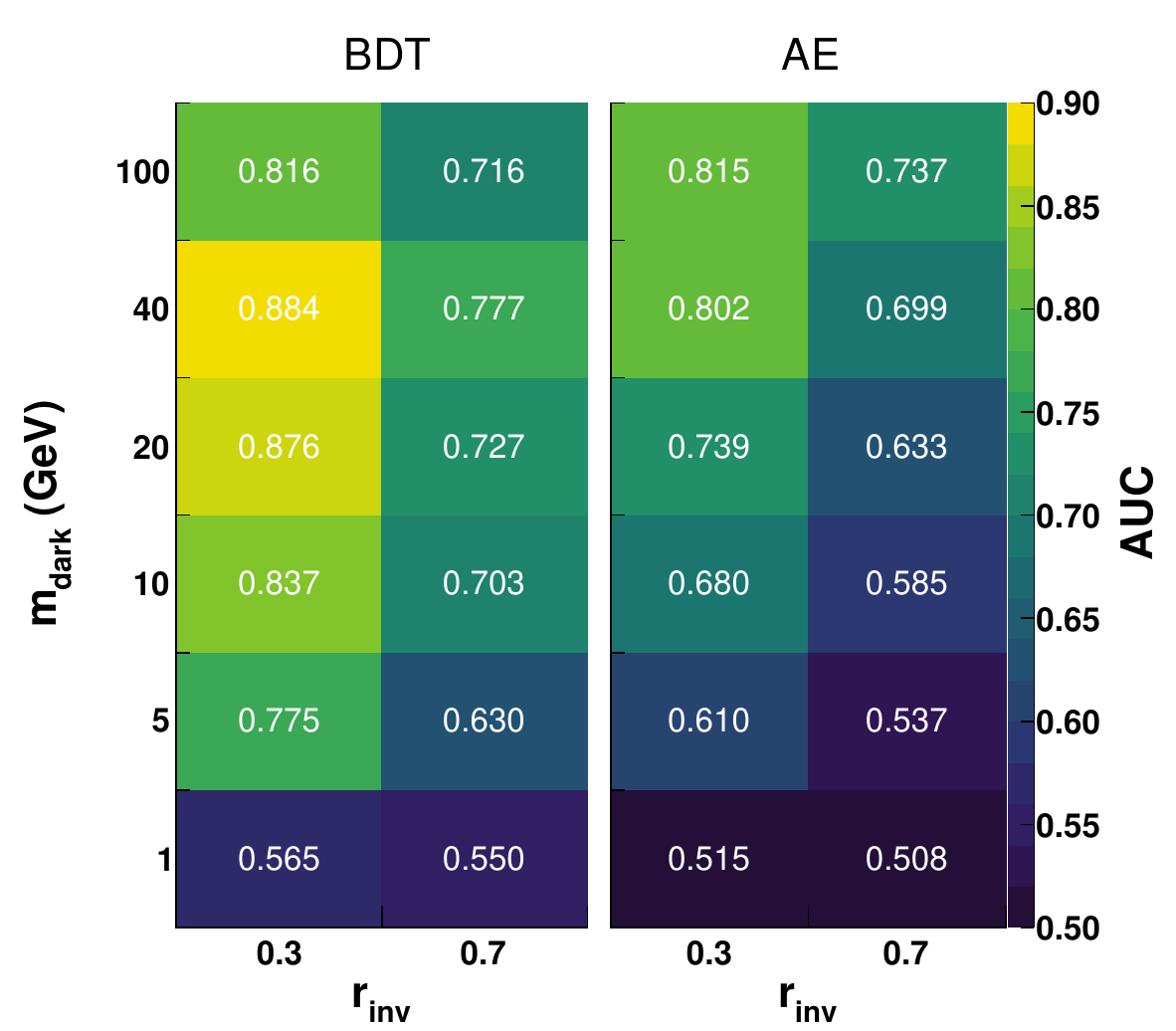}
\caption{Comparison of AUC values of the autoencoder and BDT with varying \mDark values. The BDT was trained on a mixture of all signals with $\mDark = 20$~GeV.}
\label{fig:aucs_mdark}
\end{figure}

Another robustness test was to evaluate both the BDT and the AE on signal samples with varying \mDark values, with \mZprime fixed to 3~TeV. The results of this study are presented in Fig.\ref{fig:aucs_mdark}. Here, one can see again that in certain cases the BDT trained on a wrong signal hypothesis can be outperformed by the AE. In this scenario, the AE becomes more powerful than a classifier for $\mDark = 100$~GeV, when the BDT training was performed on a mixture of 18 signal samples (with varying \mZprime and \rinv) with $\mDark = 20$~GeV. More details on the signal samples with varying \mDark values can be found in Appendix~\ref{app:mdark}.

We noticed that training the BDT on a specific signal point with low \rinv leads to higher efficiency for signals with higher \rinv with respect to training on a mixture of all signal points. This result comes from the fact that low \rinv jets are more similar to background jets. Such conclusions suggest that it might be a better strategy to train a classifier on low \rinv signals rather than on a mixture of low and high parameter values.

From this test, we conclude that it would be preferable for an LHC signal-specific search to have a signal-agnostic equivalent exploiting autoencoders to complement model-dependent searches, as opposed to relying on accidental generalization properties of the signal-specific selection algorithm. The synergy of the two approaches would enhance the chances of discovering a signal.

\subsection{Sensitivity to Semivisible Jet Models}

\begin{figure}[!tbp]
  \centering
  \subfloat[\rinv = 0.3]{
    \includegraphics[width=0.47\textwidth]{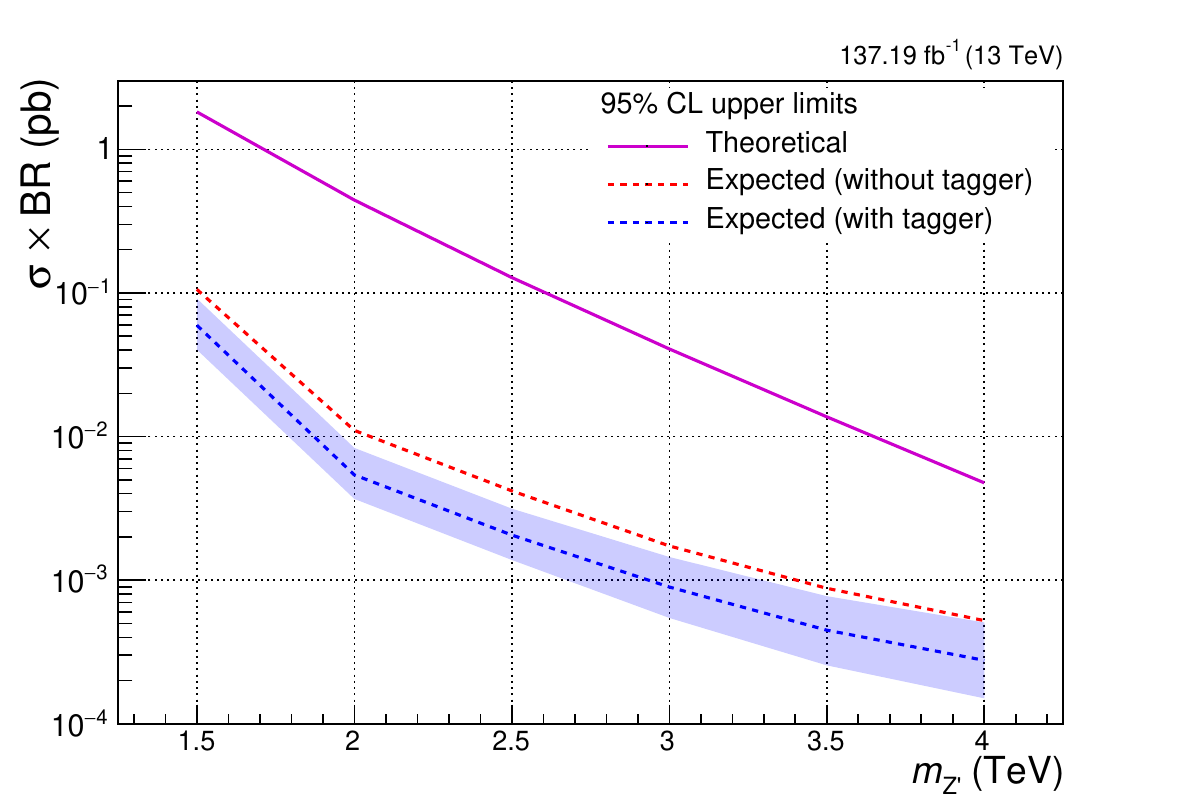}
    \label{fig:limits_0p3}
  }
  \hfill
  \subfloat[\rinv = 0.7] {
    \includegraphics[width=0.47\textwidth]{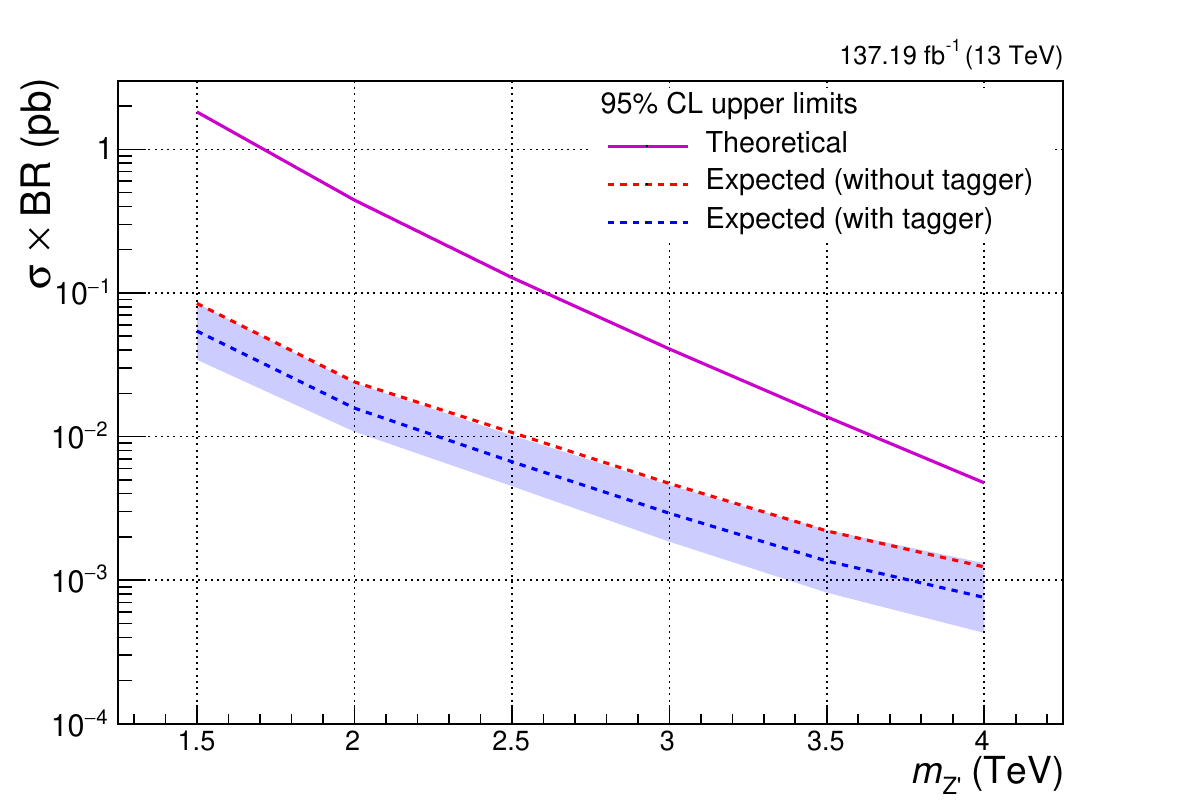}
    \label{fig:limits_0p7}
  }
  \hfill
  \caption{The cross section limits vs. \mZprime for different values of \rinv and $\mDark = 20$~GeV, scaled to the luminosity of the full Run 2 of the LHC.}
\end{figure}

As a last step, a threshold was optimized for the SVJ signal samples to tag jets with loss > 0.06 as anomalous. Doing so, one can mimic the impact of the AE algorithm on a real analysis, estimating expected upper limits. Events passing the preselection were grouped in categories with 0, 1 or 2 SV jets, and dijet transverse mass distributions for the signal and background in these categories were used to set approximate exclusion limits on the cross section, depending on the \PZprime mass (while $\mDark$ was fixed to $20$~GeV). The results, which take into account basic uncertainties (luminosity and trigger), are presented in Figs.~\ref{fig:limits_0p3} and \ref{fig:limits_0p7} and compared with the example theoretical cross section. The results without applying an autoencoder-based tagger are also shown.

As expected, the limits are stronger for lower values of \rinv and the improvement from using the tagger is more pronounced in this region. The presented results show that one can expect an approach based on the autoencoder to be able to cover a large area of the phase space.

%
\section{Conclusion}

This study shows that the usage of a neural autoencoder based on physics-motivated high-level features, despite its simplicity, is extremely effective as an anomaly detection model. This is especially true in cases of a signal with widely varying parameters, such as semivisible jets, as the model is trained on the background sample only and is therefore highly model independent. The projections of the exclusion limits for this particular class of models show a 40-60\% improvement in the search reach compared to a study without such an anomalous jet tagger. What is more, the autoencoder could be trained on the real data, accounting for any effects of an imperfect detector simulation.

This result demonstrates the usefulness of the autoencoder network and jet substructure variables such as Energy Flow Polynomials, Energy Correlation Functions, the jet \pt dispersion \ptD, and the jet ellipse axes for the detection of anomalous jets. It is also worthwhile to acknowledge certain aspects that should be further investigated in the future.

The assumption that the network should be insensitive to detector effects could be tested by simulating such issues and contaminating the training sample with affected events. Next, in order to train autoencoder on the collision data, one has to verify that a small admixture of signal events in the training sample does not influence the performance. Finally, one could try to further exploit the four-momenta of the jet constituents using convolutional or graph autoencoders, which should preserve information about the spatial distribution of particles within the jet.

\section*{Acknowledgements}

M.~Pierini is supported by the European Research Council (ERC) under the European Union's Horizon 2020 research and innovation program (Grant Agreement No. 772369). A.~de~Cosa and J.~Niedziela are supported by the Swiss National Science Fundation (SNFS) under the SNSF Eccellenza program (PCEFP2\_186878). L.~Le~Pottier is supported by the  Swiss State Secretariat for Education, Research and Innovation (SERI) under the ThinkSwiss program. K. Pedro is supported by the Fermi National Accelerator Laboratory, managed and operated by Fermi Research Alliance, LLC under Contract No. DE-AC02-07CH11359 with the U.S. Department of Energy.

\bibliography{autoencodeSVJ}

\begin{thebibliography}{10}

\bibitem{hidden_valley}
Matthew~J. Strassler and Kathryn~M. Zurek.
\newblock Echoes of a hidden valley at hadron colliders.
\newblock {\em Phys. Lett. B}, 651:374--379, 2007.

\bibitem{sv_jets}
Timothy Cohen, Mariangela Lisanti, and Hou~Keong Lou.
\newblock Semivisible jets: Dark matter undercover at the {LHC}.
\newblock {\em Phys. Rev. Lett.}, 115(17):171804, 2015.

\bibitem{Cohen:2017pzm}
Timothy Cohen, Mariangela Lisanti, Hou~Keong Lou, and Siddharth Mishra-Sharma.
\newblock {LHC Searches for Dark Sector Showers}.
\newblock {\em JHEP}, 11:196, 2017.

\bibitem{cmscollaboration2021search}
CMS Collaboration.
\newblock Search for resonant production of strongly coupled dark matter in
  proton-proton collisions at 13 {TeV}, 2021.

\bibitem{qcd_or_what}
Theo Heimel, Gregor Kasieczka, Tilman Plehn, and Jennifer~M. Thompson.
\newblock {QCD} or what?
\newblock {\em SciPost Phys.}, 6(3):030, 2019.

\bibitem{new_physics_deep_autoencoders}
Marco Farina, Yuichiro Nakai, and David Shih.
\newblock Searching for new physics with deep autoencoders.
\newblock {\em Phys. Rev. D}, 101(7):075021, 2020.

\bibitem{lhc_olympics}
Gregor Kasieczka et~al.
\newblock {The LHC Olympics 2020: A Community Challenge for Anomaly Detection
  in High Energy Physics}, 1 2021.

\bibitem{dark_machines}
T.~Aarrestad et~al.
\newblock {The Dark Machines Anomaly Score Challenge: Benchmark Data and Model
  Independent Event Classification for the Large Hadron Collider}, 5 2021.

\bibitem{cheng2021variational}
Taoli Cheng, Jean-François Arguin, Julien Leissner-Martin, Jacinthe Pilette,
  and Tobias Golling.
\newblock Variational autoencoders for anomalous jet tagging, 2021.

\bibitem{olmo}
Olmo Cerri, Thong~Q. Nguyen, Maurizio Pierini, Maria Spiropulu, and Jean-Roch
  Vlimant.
\newblock Variational {Autoencoders} for {New} {Physics} {Mining} at the
  {Large} {Hadron} {Collider}.
\newblock {\em Journal of High Energy Physics}, 2019(5):36, May 2019.
\newblock arXiv: 1811.10276.

\bibitem{eflow}
Patrick~T. Komiske, Eric~M. Metodiev, and Jesse Thaler.
\newblock Energy flow polynomials: A complete linear basis for jet
  substructure.
\newblock {\em JHEP}, 04:013, 2018.

\bibitem{ecf}
Andrew~J. Larkoski, Gavin~P. Salam, and Jesse Thaler.
\newblock Energy correlation functions for jet substructure.
\newblock {\em JHEP}, 06:108, 2013.

\bibitem{ptd}
{CMS Collaboration}.
\newblock {Performance of quark/gluon discrimination in 8 TeV pp data}.
\newblock {CMS Physics Analysis Summary}, CERN, Geneva, 2013.

\bibitem{jets_at_lhc}
Roman Kogler et~al.
\newblock {Jet Substructure at the Large Hadron Collider: Experimental Review}.
\newblock {\em Rev. Mod. Phys.}, 91(4):045003, 2019.

\bibitem{svj_substructure}
Deepak Kar and Sukanya Sinha.
\newblock Exploring jet substructure in semi-visible jets.
\newblock {\em SciPost Physics}, 10(4), Apr 2021.

\bibitem{emerging_jets}
Pedro Schwaller, Daniel Stolarski, and Andreas Weiler.
\newblock Emerging jets.
\newblock {\em Journal of High Energy Physics}, 2015(5), May 2015.

\bibitem{graph_net_dark_showers}
Elias Bernreuther, Thorben Finke, Felix Kahlhoefer, Michael Krämer, and
  Alexander Mück.
\newblock Casting a graph net to catch dark showers.
\newblock {\em SciPost Physics}, 10(2), Feb 2021.

\bibitem{parametric_ml}
Pierre Baldi, Kyle Cranmer, Taylor Faucett, Peter Sadowski, and Daniel
  Whiteson.
\newblock Parameterized neural networks for high-energy physics.
\newblock {\em The European Physical Journal C}, 76(5), Apr 2016.

\bibitem{pythia}
Torbj\"orn Sj\"ostrand, Stefan Ask, Jesper~R. Christiansen, Richard Corke,
  Nishita Desai, Philip Ilten, Stephen Mrenna, Stefan Prestel, Christine~O.
  Rasmussen, and Peter~Z. Skands.
\newblock An introduction to {PYTHIA} 8.2.
\newblock {\em Comput. Phys. Commun.}, 191:159--177, 2015.

\bibitem{hepdata.115426}
{CMS Collaboration}.
\newblock Search for resonant production of strongly coupled dark matter in
  proton-proton collisions at 13 {TeV}.
\newblock {HEPData (collection)}, 2021.
\newblock \url{https://doi.org/10.17182/hepdata.115426}.

\bibitem{delphes}
J.~de~Favereau, C.~Delaere, P.~Demin, A.~Giammanco, V.~Lema\^\i{}tre,
  A.~Mertens, and M.~Selvaggi.
\newblock {DELPHES} 3, a modular framework for fast simulation of a generic
  collider experiment.
\newblock {\em JHEP}, 02:057, 2014.

\bibitem{jet_substructure}
Jason Gallicchio and Matthew~D. Schwartz.
\newblock Quark and gluon jet substructure.
\newblock {\em JHEP}, 04:090, 2013.

\bibitem{lha}
J.~R. Andersen et~al.
\newblock {Les Houches 2015: Physics at TeV Colliders Standard Model Working
  Group Report}.
\newblock In {\em {9th Les Houches Workshop on Physics at TeV Colliders}}, 5
  2016.

\bibitem{ae_anomaly_hep}
Thorben Finke, Michael Kr\"amer, Alessandro Morandini, Alexander M\"uck, and
  Ivan Oleksiyuk.
\newblock {Autoencoders for unsupervised anomaly detection in high energy
  physics}.
\newblock {\em JHEP}, 06:161, 2021.

\bibitem{latent_space_classifiers}
Barry Dillon, Tilman Plehn, Christof Sauer, and Peter Sorrenson.
\newblock Better latent spaces for better autoencoders.
\newblock {\em SciPost Physics}, 11(3), Sep 2021.

\bibitem{vae}
Diederik~P. Kingma and Max Welling.
\newblock Auto-encoding variational {Bayes}.
\newblock 12 2013.

\bibitem{pca}
Ian Jolliffe.
\newblock {\em Principal Component Analysis}, pages 1094--1096.
\newblock Springer Berlin Heidelberg, Berlin, Heidelberg, 2011.

\bibitem{sae:eth}
Andrea Borghesi, Andrea Bartolini, Michele Lombardi, Michela Milano, and Luca
  Benini.
\newblock Anomaly {Detection} using {Autoencoders} in {High} {Performance}
  {Computing} {Systems}.
\newblock {\em arXiv:1811.05269 [cs]}, November 2018.
\newblock arXiv: 1811.05269.

\bibitem{elu}
Djork{-}Arn{\'{e}} Clevert, Thomas Unterthiner, and Sepp Hochreiter.
\newblock Fast and accurate deep network learning by exponential linear units
  {(ELUs)}.
\newblock In {\em {4th International Conference on Learning Representations,
  {ICLR} 2016, San Juan, Puerto Rico, May 2-4, 2016, Conference Track
  Proceedings}}, 2016.

\bibitem{keras}
Fran\c{c}ois Chollet et~al.
\newblock Keras.
\newblock \url{https://github.com/fchollet/keras}, 2015.

\bibitem{tensorflow}
Mart\'{\i}n~Abadi et. al.
\newblock {TensorFlow}: Large-scale machine learning on heterogeneous systems,
  2015.
\newblock Software available from tensorflow.org.

\bibitem{nadam}
Timothy Dozat.
\newblock Incorporating nesterov momentum into adam.

\bibitem{mahalanobis_distance}
G.~Mclachlan.
\newblock Mahalanobis distance.
\newblock {\em Resonance}, 4:20--26, 06 1999.

\bibitem{kl_divergence}
James~M. Joyce.
\newblock {\em Kullback-Leibler Divergence}, pages 720--722.
\newblock Springer Berlin Heidelberg, Berlin, Heidelberg, 2011.

\end{thebibliography}

\clearpage

\appendix
\section{\texorpdfstring{\mDark}{m\_dark} Variations}\label{app:mdark}

\begin{figure}[h]
    \centering
    \includegraphics[width=1.0\textwidth]{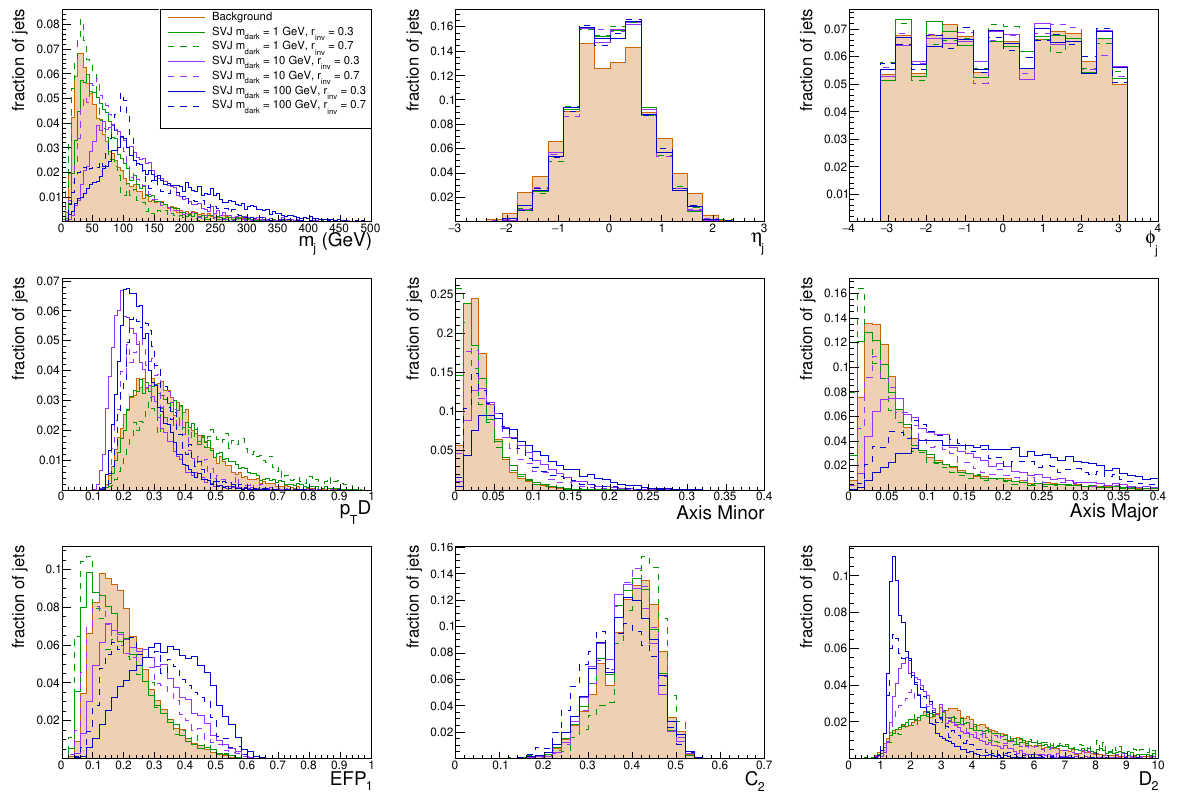}
    \caption{Distributions of input variables for QCD background and signal models with \mZprime = 3 TeV and varying \mDark and \rinv values.}
    \label{fig:inputs_mdark}
\end{figure}

\begin{figure}\centering
\includegraphics[width=1.0\linewidth]{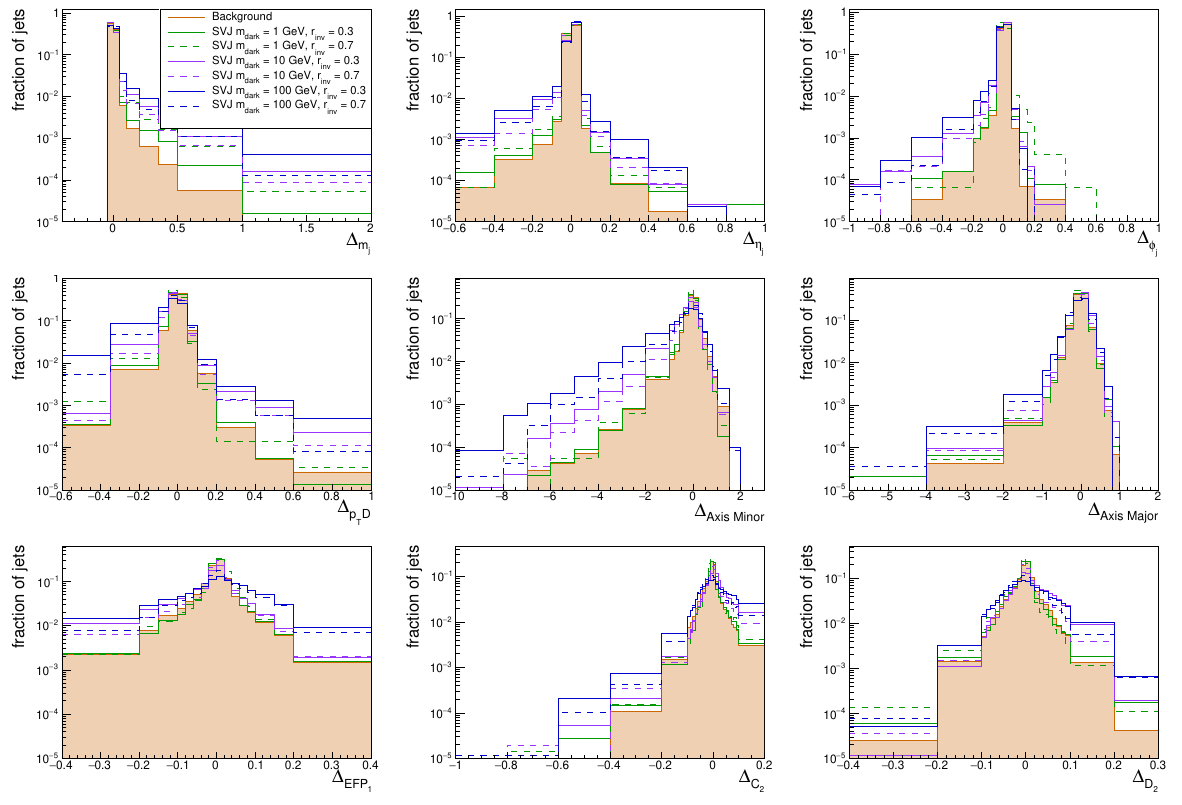}
\caption{Distributions of difference between input and reconstructed jet features for QCD and signal models at \mZprime = 3 TeV and varying \mDark and \rinv values.}
\label{fig:input_and_reco_mdark}
\end{figure}

Figure~\ref{fig:inputs_mdark} displays the input variables used in training the tagger algorithms for signal samples with different \mDark values. In particular, the effect of a large \mDark value can be seen in the \mj distribution. Figure~\ref{fig:input_and_reco_mdark} displays the reconstruction error for these variables and signal samples after applying the AE. The signal distributions show only moderate differences over a large range of \mDark values. These figures can be compared, respectively, to Figs.~\ref{fig:inputs} and \ref{fig:input_and_reco}, which show the variations in \rinv and \mZprime.

\end{document}